# Ionic environment control of visible photo-luminescence from ZnO nanoparticles

Manoranjan Ghosh and A.K.Raychaudhuri

*DST Unit for Nanoscience, S.N.Bose National Centre for Basic Sciences,*

*Block-JD, Sector-3, Salt Lake, Kolkata-700 098, INDIA*

We report a novel effect that the visible photoluminescence (in the blue-green band) from ZnO nanoparticles can be controlled by changing the ionic or polar nature of the medium in which the nanoparticles are dispersed. We find that presence of sufficient amount of electrolytes can even quench the emission. We propose an explanation based on surface charge of the ZnO nanoparticles which control the band bending in the depletion layer at the surface of the nanoparticles. The band bending in turn, decides the predominant nature of the visible emission. The explanation is validated by establishing a direct correlation between the visible emission and the zeta potential.

**Keywords:** ZnO nanostructure, visible emission, Oxygen vacancy, Zeta Potential,

Recently, ZnO has emerged as one of the most researched wide band gap (3.3 eV) optoelectronic material due to its attractive optical properties. It is a cheap, non-toxic material and can be easily synthesized. Bulk ZnO crystal shows a sharp emission in the ultraviolet region due to excitonic recombination (exciton binding energy~ 60 meV) which makes ZnO a promising material for UV laser and light emitting diode (LED) at room temperature [1]. Interestingly, ZnO nanostructures also show a broad visible photoluminescence (PL) in the blue green region (470 nm–550 nm) [2]. The visible emission originates from defects such as Oxygen vacancy which are believed to be located near the surface region [2,3]. It has been observed that the emission energy as well as the intensity of the visible photoluminescence not only depends on the size, but is also related to the morphology of the nanostructures [4]. For example, spherical ZnO nanoparticles of size less than 20 nm show broad visible photoluminescence around 550 nm but the nanorods (of diameter > 20 nm and length ~ 100-150 nm) show a peak around 500 nm [4]. In this paper we report a new physical phenomenon where the visible emission from ZnO nanostructure was controlled (including near complete quenching) by controlling the ionic environment of the medium in which the nanostructures were suspended. By zeta potential measurement, we could establish a direct correlation between the surface potential of the nanostructures (nanoparticles and nanorods) and the visible emission. This correlation enabled us to propose an explanation of the observed phenomenon, which is interesting because of its application potential.

In this study, two kinds of nanostructures have been chosen: the spherical particles in which 550 nm (2.2 eV) emission band dominates and the cylindrical nanorods whose visible emission is dominated by the 500 nm (2.5 eV) band. The spherical particles [figure 1 (a)] with typical average diameter of 10nm have been synthesized by acetate route under ambient pressure [5]. The nanorods of dia ~ 20 nm and length 100-150nm [figure 1 (b)] were prepared in an autoclave by the same chemical route but under 54 atm. pressure at 230 $^0$C [6]. The nanostructures were imaged with a JEOL high-resolution transmission electron microscope (TEM) at 200 KeV. X-ray diffraction data confirms the wurtzite symmetry of the synthesized nanostructures. The TEM images along with high resolution microscopy data are shown in figure 1 for both the nanostructures. The materials synthesized have also been characterized by X-ray-diffraction. The average size calculated from the X-ray diffraction data matches well with that found from TEM micrograph.

Below we first describe the basic phenomena as observed. In this experiment we measure the PL spectra of the nanostructures dispersed in a solution with different strengths of the electrolytes such as LiClO$_4$ (in ethanol) and NaCl (in water). Prior to taking the PL data the absorbance data were taken and the fundamental absorption edge was found to be at 3.566 eV (348 nm) for the particles and 3.371 eV (368 nm) for the nanorods. The fundamental absorption edge of the nanorods remains unchanged in the presence of electrolytes, however, a small red shift of 0.07 eV (for 1 M LiClO$_4$) has been observed in case of nanospheres. Room temperature optical properties were studied by photoluminescence spectrometer (Jobin Yvon, FluoroMax-3) using a xenon arc lamp as illumination source at an excitation wavelength of 325 nm.

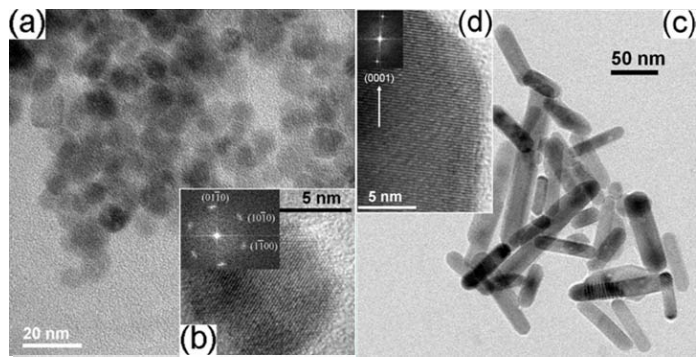

Figure 1: TEM micrographs for the (a) nanosphere and (c) nanorods of ZnO. The FFT of the HRTEM of a nanosphere shown in (b) confirms the hexagonal symmetry. Nanorods growth in the [0001] direction is shown (d) by the lattice image and its FFT of a single rod.

In figure 2 (a), we plot PL spectra of spherical ZnO nanoparticles dispersed in ethanol having different concentrations of LiClO$_4$ as indicated in the graph. It can be clearly seen that as the concentration of the electrolyte increases there is a significant change in the emission spectra. In fact for higher concentration of the electrolyte there is almost complete quenching of the visible emission. Maximum 70 % decrease in total intensity on changing the electrolyte concentration can be seen. In the inset of figure 2 (a) we show the integrated intensity of the visible emission as a function of ionic concentration. Similar effects, although with much reduced magnitude has also been observed for the nanorods. The data is plotted in figure 3(a). The inset shows dependence of the integrated emission intensity as a function of the electrolyte concentration. The observed data, both for the nanoparticles and the nanorods, show unambiguously that the intensity of the visible emission from ZnO nanostructures can be very effectively controlled by the ionic nature of the environment in which they are suspended. To our knowledge such an effect of the electrolyte (or the ionic nature of the medium) on the visible emission of ZnO nanostructures has not been reported earlier.

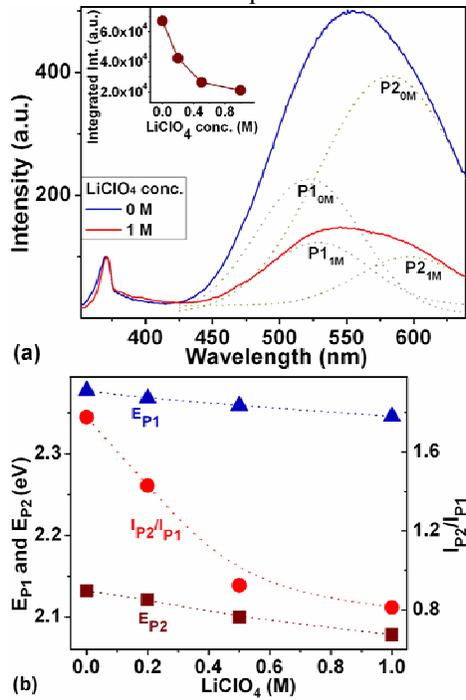

Figure 2: (a) Photoluminescence spectra of ZnO nanospheres for different $LiClO_4$ concentrations in ethanol as indicated in the graph. Inset shows the integrated intensity with different concentration of electrolyte. (b) Emission energies and the relative intensity ratios of the spectral components have been plotted

The visible emission is also sensitive to the polar nature of the medium of suspension. For example, the visible emission intensity is higher when ZnO is suspended in water compared to that when it is suspended in less polar ethanol. Interestingly, even addition of NaCl in water suppresses the visible emission.

Below we carry out a quantitative evaluation of the observed changes and offer an explanation for the observed phenomena. It has been known before that the visible emission in the blue-green region, as we see here is a composite of two broad lines located approximately at 2.2 eV (550nm) and at 2.5 eV (500nm) [7]. The spectral nature of the visible emission band in this region depends crucially on the exact position as well as the relative emission intensities from these two lines. The nature of the visible emission in the spherical particles as well as the nanorods changes because of the relative strengths of the two emission lines. In figure 2 (a) we show the two decomposed peaks of the broad visible emission for the particles dispersed in pure ethanol (marked $P1_{0M}$ and $P2_{0M}$) and 1M $LiClO_4$ solution in ethanol (marked $P1_{1M}$ and $P2_{1M}$). Similar decomposition into the two lines is shown for the emission from the nanorods in figure 3(a). The important thing to note is that the relative strength of the two lines changes as the electrolyte concentration changes. In figure 2 (b) as well as 3(b) we plot the intensity ratio ($I_{P2}/I_{P1}$) between the two spectral components P2 and P1). The change in the emission spectra originates predominantly due to a reduction in the relative contribution of the P2 line on increasing the $LiClO_4$ concentration. (The ratio $I_{P2}/I_{P1}$ decreases monotonously for both types of nanostructures, although it is most prominent in the smaller nanoparticles.) The spectral positions of the two components show a small but gradual shift on increasing the electrolyte concentration. In case of the nanoparticles for maximum concentration of $LiClO_4$, the observed shift in the spectral position of P2 ($E_{P2}$) is more (0.054 eV) in comparison to that of P1 (0.032 eV). This shows that the P2 emission is more sensitive to the ionic environment. An important point need be noted. The near band edge emission energy (at ~380nm, related to excitons) is not affected at all by the presence of the electrolyte.

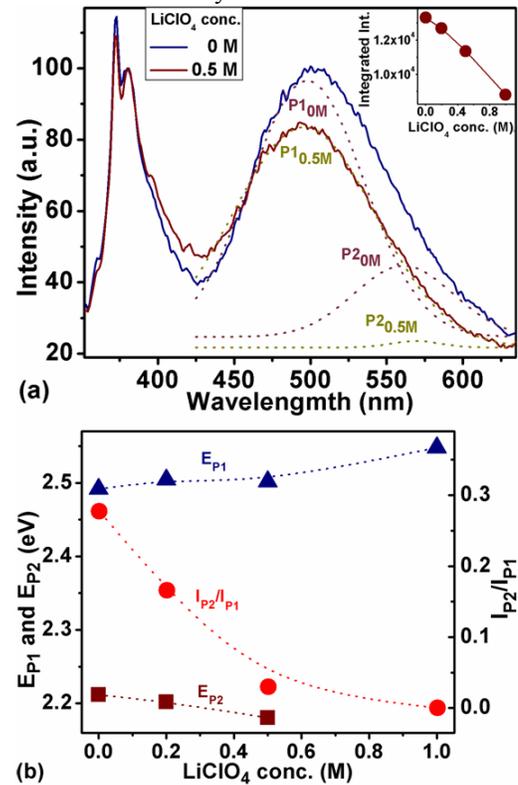

Figure 3.(a) Photoluminescence spectra of ZnO nanorods showing the contributions of the two spectral components for 0 M and 0.2 M $LiClO_4$ in ethanol. The inset shows the integrated intensity for different electrolyte concentration. (b) The emission energies and the relative intensity ratios are plotted with different concentrations of electrolyte.

As stated in the nanorods of relatively larger size (dia~20 nm) dispersed in ethanol, there is a rapid fall in P2 emission on addition of $LiClO_4$ to the solution [figure 3] and P1 emerges as the dominant component. For the maximum concentration of $LiClO_4$ (1 M), P2 does not have any contribution at all. Thus the nanorods behave similarly as in the case of nanoparticles and show P1 as the dominant emission band which remains almost unchanged.

In the remaining part of the paper we discuss the likely origin of the observation. We first propose the postulate that the change in the visible emission (or the change in the components P1 and P2) occur due to change in the surface charge or surface potential of the nanoparticles or nanorods. We prove correctness of this postulate by actually measuring the zeta potential ($\xi$) of the nanostructures and correlating with the intensity of the P2 line. The Zeta potentials were measured at $20^0$C using Malvern Instruments' Zetasizer systems Nano ZS.

The value of $\xi$ for the ZnO nanostructures in different media first decreases rapidly and then slowly reduces to a constant value as we increase the electrolyte concentration. Even in pure ethanol, the small spherical particles of size ~ 10 nm show higher $\xi$ value ($\approx$18mV) at pH 7.3 than the nanorods of relatively larger diameter ($\xi \approx$ 8 mV at pH 7). This reduction in positive $\xi$ thus leads to a reduction in the visible emission and in particular the reduction of the ratio $I_{P2}/I_{P1}$. Evidently, the increase in electrolyte concentration, reduce $\xi$ in both nanoparticles and nanorods significantly. The strong correlation of the $\xi$ with the emission intensity can be seen in figure 5 where we have plotted ratio $I_{P2}/I_{P1}$ as a function of $\xi$. Decrease in $\xi$ reduces the relative strength of the P2 line. This is definitely a very important observation that the spectral control of the relative strengths of the two lines can be done by controlling the $\xi$.

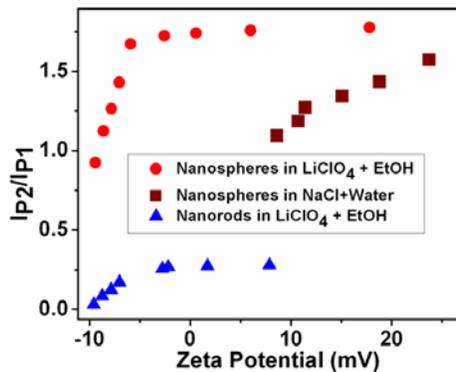

Figure 4. The intensity ratios of P2 and P1 bands ($I_{P2}/I_{P1}$) are plotted as a function of Zeta potential for both the nanostructures suspended in different media as indicated on the graph.

It has been established that the visible emission from ZnO in the blue-green region originates from defects (oxygen vacancy in particular) which are located predominantly near the surface of the nanoparticles [2,3]. Emission band appearing around 550 nm (P2) has been suggested to originate from doubly charged oxygen vacancy ($V_o^{++}$) whereas singly charged oxygen vacancy ($V_o^{+}$) is responsible for the emission band around 500 nm (P1) [7]. The relative intensities of the two lines will depend on their relative occupancy. Native ZnO nanoparticles are n-type. But there exists a depletion layer near the surface [7]. This makes the surface of ZnO nanoparticles positively charged. Such a depletion layer gives rise to band bending in spherical nanoparticles whose size is comparable to depletion width (~5-10 nm) [8]. For positive charge on the surface, the band bending lifts the chemical potential which populates the level $V_o^{++}$ preferentially compared to the other level, leading to a relative enhancement of the intensity of the line P2. This has been observed in the past and it establishes the essential roles of the surface charge and the band bending in determination of the spectral components of the blue green visible emission from ZnO nanoparticles [7]. The band bending in the near surface depletion layer in the nanoparticles is linked to their surface charge. So any modification of the surface charge will be reflected by change in the level of band bending. This will be more predominant when the size of the nanoparticles decreases. The direct correlation between the visible emission and the zeta-potential thus leads to the explanation that the change in the visible emission in an electrolytic medium occurs due to change in the band bending which in turn is caused by a change in the effective surface charge of the ZnO nanoparticles. When $\xi$ decreases due to reduced positive charge on the nanoparticles surface, the band bending is reduced and the component P2 arising from the $V_o^{++}$ state, makes diminishing contribution.

The above explanation of the quenching of the visible emission opens up a way to control the visible emission form the ZnO nanoparticles. In particular, it can be used as a visible bio-sensor for biochemical/biophysical processes that leads to a change in the surface charge and the zeta-potential. We note that this effect will depend on size of particles and will be enhanced in smaller particles whose size will be comparable to or smaller than the depletion width. In larger particle where predominant contribution arises from the level $V_o^{+}$ (line P1), the effect of band bending will be reduced and correspondingly the effect of the ionic environment is also reduced.

To summarize, we have shown that visible photoemission from small nanoparticles of ZnO can be effectively controlled and even quenched by simply varying the ionic content or polar nature of the suspending medium. An explanation has been proposed that depends on control of the predominant emission centres ($V_o^{++}$ or $V_0^{+}$) by controlling the band bending in the depletion layer near the surface which in turn can be controlled by the surface charge.

The authors acknowledge the financial support from the Department of Science and Technology, Government of India as a Unit for Nanoscience. The authors would also like to thank to M. Venkat Kamlakar for useful discussions, IACS for providing TEM facility and IUCCSR, Kolkata for Zeta potential measurement.